# Tuning spin-orbit torques across the phase transition in VO₂/NiFe heterostructure

Jun-young KIM[1,2], Joel CRAMER[1], Kyujoon LEE[1,3], Dong-Soo HAN[1,4], Dongwook GO[1,5], Pavel SALEV[6], Pavel N. LAPA[6], Nicolas M. VARGAS[6], Ivan K. SCHULLER[6], Yuriy MOKROUSOV[1,5], Gerhard JAKOB[1], and Mathias KLÄUI[1]

[1]Institute of Physics, Johannes Gutenberg University, 55128 Mainz, Germany

[2]Max Planck Institute for Intelligent Systems, 70569 Stuttgart, Germany

[3]Department of Semiconductor Physics, Korea University, 30019 Sejong, Republic of Korea

[4]Korea Institute of Science and Technology, 02792 Seoul, Republic of Korea

[5]Peter Grünberg Institut and Institute for Advanced Simulation, Forschungszentrum Jülich and JARA, 52425 Jülich, Germany

[6]Department of Physics and Center for Advanced Nanoscience, University of California San Diego, 92093 La Jolla, CA, USA

## ABSTRACT

The emergence of spin-orbit torques as a promising approach to energy-efficient magnetic switching has generated large interest in material systems with easily and fully tunable spin-orbit torques. Here, current-induced spin-orbit torques in VO₂/NiFe heterostructures were investigated using spin-torque ferromagnetic resonance, where the VO₂ layer undergoes a prominent insulator-metal transition. A roughly two-fold increase in the Gilbert damping parameter, $\alpha$, with temperature was attributed to the change in the VO₂/NiFe interface spin absorption across the VO₂ phase transition. More remarkably, a large modulation ($\pm$ 100%) and a sign change of the current-induced spin-orbit torque across the VO₂ phase transition suggest two competing spin-orbit torque generating mechanisms. The bulk spin Hall effect in metallic VO₂, corroborated by our first-principles calculation of spin Hall conductivity $\sigma_{\text{SH}} \sim -10^4 \left(\frac{\hbar}{e}\right)\Omega^{-1}\text{m}^{-1}$, is verified as the main source of the spin-orbit torque in the metallic phase. The self-induced/anomalous torque in NiFe, of the opposite sign and a similar magnitude to the bulk spin Hall effect in metallic $VO_2$, could be the other competing mechanism that dominates as temperature decreases. For applications, the strong tunability of the torque strength and direction opens a new route to tailor spin-orbit torques of materials which undergo phase transitions for new device functionalities.

---

[1]Correspondence to: klaeui@uni-mainz.de

* Author to whom correspondence should be addressed: klaeui@uni-mainz.de

## 1. INTRODUCTION

Long-term goals of spintronics are the generation and the utilisation of spin currents for information processing and storage[1,2]. Compared to optical and electrical spin injection schemes[3,4], the spin current generation via spin-orbit interaction has demonstrated efficient charge-to-spin conversion[5,6] and has received much interest not only in the fundamental understanding but also in particular for technological applications. Notably, magnetisation switching via spin-orbit torques[7,8] offers a number of advantages over conventional spin-transfer torque switching and is actively being developed into new generation spintronics devices such as spin-orbit torque magnetoresistance random access memory[9].

The main mechanisms behind these recent advances are the current-induced spin-orbit torques[10,11]. The torques can be realised in a number of different ferromagnet-nonmagnet systems, and efficient charge-to-spin conversion was observed not only in conventional metallic heterostructures but also in non-magnetic metal bilayers[12], semiconductor quantum wells[13–15] and topological insulators[16–18]. So far various mechanisms for the observed spin-orbit torques have been identified, however, very often it has been challenging to identify the origin of the spin-orbit torques because different mechanisms contribute at the same time and compete with each other. Furthermore, varying layer thicknesses in order to disentangle bulk and interface effects poses difficulties as the growth and interface properties change with varying the thicknesses. For the bulk effects, the spin Hall effect of the nonmagnet has been regarded as one of the main contributions and the values can now be calculated also theoretically[6,8]. Meanwhile, while the effect of spin-orbit coupling in the ferromagnet has been regarded negligible so far, a recent experiment[19] revealed that even a single ferromagnet can generate substantial self-induced torque with a defined sign of the torque. Moreover, it was shown that orbital Hall current generated from the nonmagnetic layer can also contribute strongly to the torque[20].

The spin-orbit torque efficiency is a parameter that is usually set for a specific material and interface, and it cannot be modulated easily. While it has been recently shown that strain can be used to control the spin-orbit torque to some extent[21], a piezoelectric substrate is often required which complicates growth and optimisation of thin films. In this respect, an interesting material is vanadium dioxide ($VO_2$), a transition metal oxide which undergoes a prominent insulator-metal transition with temperature. The hysteretic phase transition allows to deliberately switch between insulating and metallic states, which can then influence the current flow and thus spin-orbit effects. The change in the $VO_2$ orbital occupation[22] across the structural phase transition leads to the large changes in electrical resistivity[23] as well as optical[24], structural[25] and magnetic properties[26–28], and is expected to affect the spin-orbit coupling directly[29]. However, the effect of the $VO_2$ phase transition on current-induced spin-orbit torques in a $VO_2$/ferromagnet heterostructure, which is of the key importance for the future functionalisation, has not been investigated and therefore is the main focus of this study.

In this work, we investigate current-induced spin-orbit torques in a $VO_2$/NiFe heterostructure across the $VO_2$ insulator-metal phase transition with the emphasis on the functionalisation. The sign and the magnitude of the generated spin-orbit torques are probed using the spin-torque ferromagnetic resonance (ST-FMR) technique, where we inspect resonance linewidths of the bilayer strips with an additional DC current through the strip. Due to the several orders of magnitude changes in the electrical resistivity of the $VO_2$ layer across the phase transition, the ratio of applied charge currents in $VO_2$ and NiFe layers is thus controlled by changing temperature. In particular, we quantify the large variation including a sign change of spin-orbit torque in different $VO_2$ phases. The observed hysteretic, phase-dependent spin-orbit torques could be utilised for future device concepts.

## 2. ST-FMR MEASUREMENTS ACROSS INSULATOR-METAL TRANSITION

Firstly, several structural characterisations were performed to inspect the quality of the $VO_2$ films. Figure 1a shows an out-of-plane x-ray diffraction spectrum of the 70 nm thick $VO_2$ film deposited by reactive sputtering on $Al_2O_3$(1-102). The (110), (200), and (111) $VO_2$ peaks are visible, indicating the polycrystalline growth of the $VO_2$ film. In Figure 1b, a 1 μm x 1 μm atomic force microscope image of the same film shows large structural domains of a few hundred nm sizes with a root-mean-square roughness of 6.3 nm. Figure 1c displays the temperature dependence of van der Pauw resistance of the as-deposited $VO_2$ film, where an insulator-metal transition with temperature yields with a resistance change of four orders of magnitude, as observed previously[23]. In order to characterise the current-induced spin-orbit torques of the $VO_2$ layers, a $Ni_{81}Fe_{19}$ (5 nm) / MgO (2 nm) / Ta (3 nm) multilayer stack is sputter-deposited on top of the VO₂ (70 nm) film. *M-H* hysteresis loops of the resulting multilayer stack measured at 300 K are shown in Figure 1d. The magnetically soft NiFe layer is fully saturated by a 10 mT magnetic field to within 10% of the expected bulk saturation value of $\sim 8.8 \times 10^5$ A m$^{-1}$, with the coercivity below 1 mT.

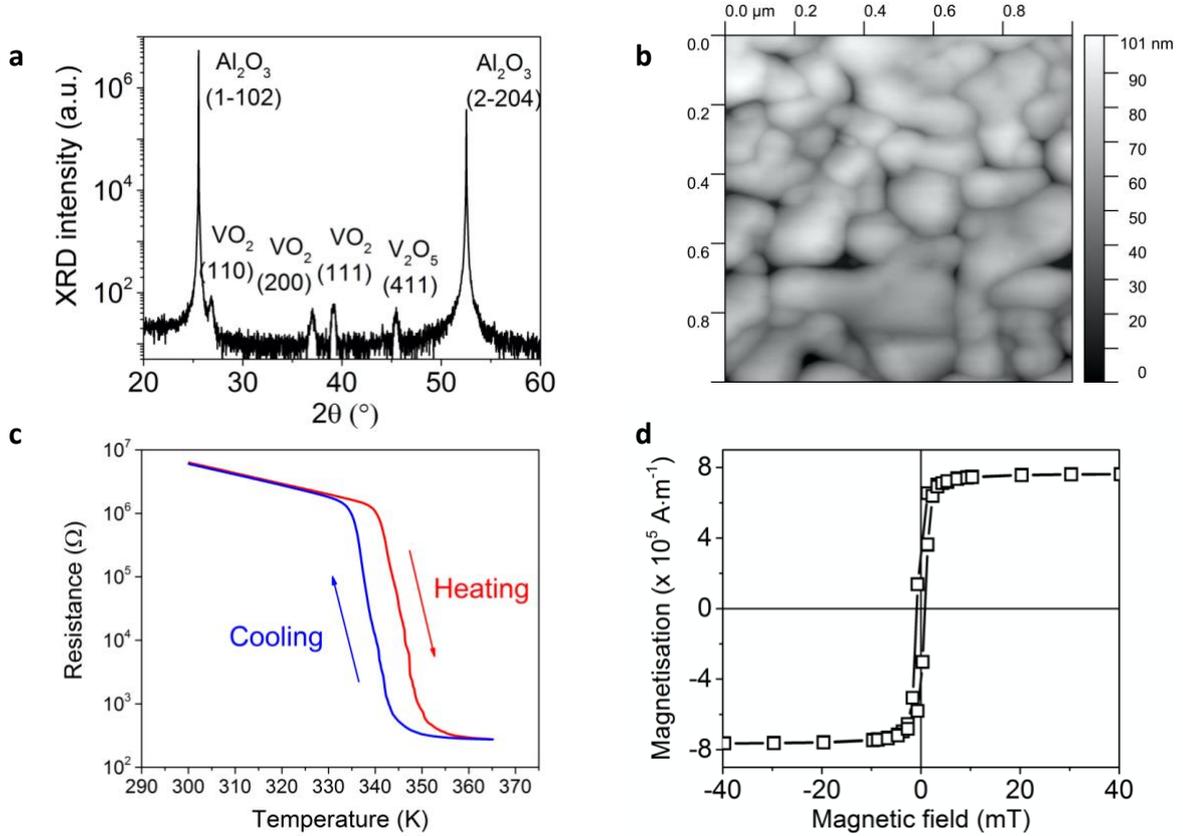

**Fig. 1: Structural properties and insulator-metal transition of VO₂. a,** Out-of-plane XRD scan of the 70 nm VO₂ film on the Al₂O₃(1-102) substrate. **b,c,** 1 μm x 1 μm atomic force micrograph (**b**) and temperature dependence of four-probe van der Pauw resistance (**c**) of the VO₂ film. **d,** SQUID *M-H* loops of the VO₂ (70 nm) / NiFe (5 nm) bilayer measured at 300 K.

The VO₂/NiFe/MgO/Ta structure (Figure 2a) is patterned into 10 μm wide strip and embedded into a coplanar-wave-guide circuit, as seen in the measurement schematic in Figure 2b. Details of the patterning process can be found in Supporting Information (Section II). The temperature-dependent resistance of the patterned device stack is shown in Figure 2c. In the patterned strip, the insulator-metal transition becomes broader in temperature than in the unpatterned film (Figure 1c), which we account to edge defects created during the patterning process. ST-FMR field sweeps of the sample measured at different sample temperatures between 290 K and 355 K are shown in Figure 2d. The sample temperatures were controlled using a Peltier element directly under the sample. The ST-FMR signal, $V_{mix}$, which effectively measures the RF-current-rectified anisotropic magnetoresistance of the strip, is enhanced when the applied magnetic field and the RF current frequency matches the resonance condition for the NiFe moment precession. The obtained ST-FMR line shape can be represented as a superposition of symmetric and anti-symmetric Lorentzian components using the equation[30,31]

$$V_{mix} = S\frac{W^2}{(\mu_0 H - \mu_0 H_{res})^2 + W^2} + A\frac{W(\mu_0 H - \mu_0 H_{res})}{(\mu_0 H - \mu_0 H_{res})^2 + W^2} + V_{const} \qquad (1)$$

where $S$ and $A$ are the symmetric and the anti-symmetric coefficients, $W$ is the resonance linewidth, $\mu_0$ is the vacuum permeability, $H_{res}$ is the resonance field and $V_{const}$ is an offset voltage in the measurement. The symmetric component, S, of $V_{mix}$ is proportional to the damping-like torque generated by the spin current from the bulk VO₂ layer and the VO₂/NiFe interface, while the anti-symmetric component, A, is generated by the Oersted field produced by the RF excitation current as well as the field-like torque arising from the spin current. The RF frequency dependence of $H_{res}$ and $W$ can be found in Supporting Information (Figure S2). As the sample temperature increases, the VO₂ becomes more metallic and the amount of the RF current through the NiFe layer that produces the $V_{mix}$ signal decreases.

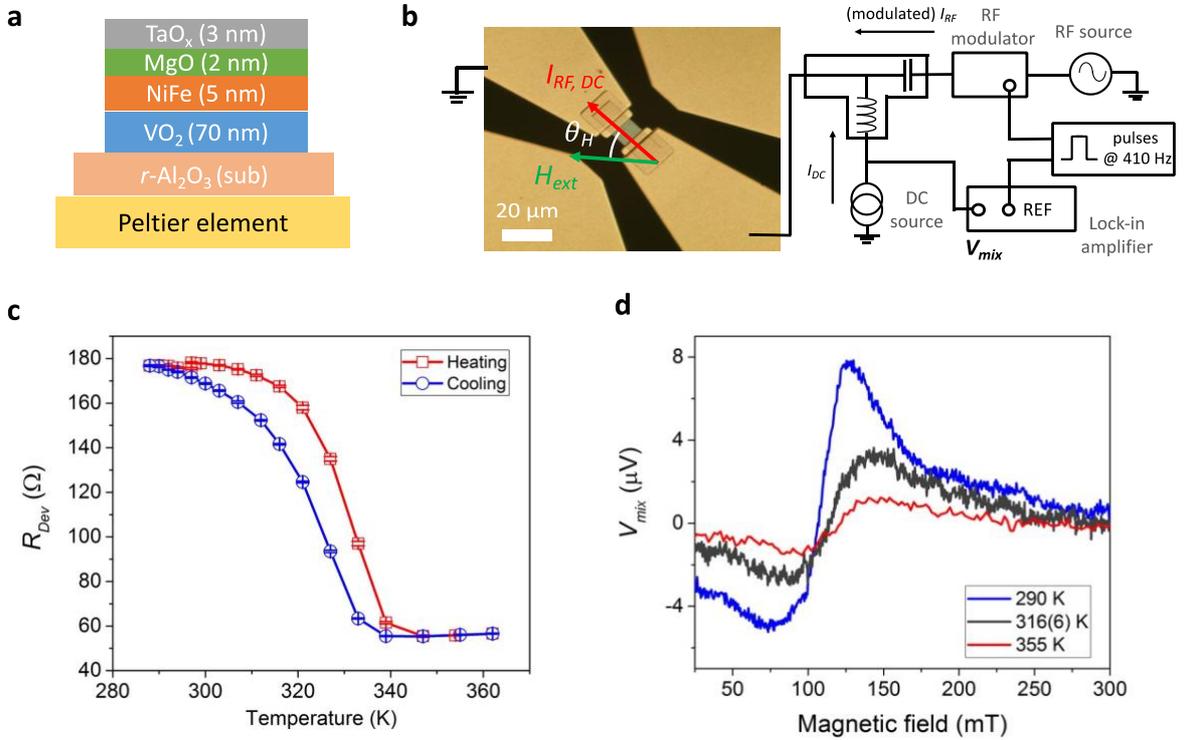

**Fig. 2: VO₂/NiFe bilayer sample and ST-FMR measurements. a,** Cross section of the studied heterostructure and microscope image of the 10 µm-wide device embedded into the CPW circuit. **b**, The experimental set-up used for ST-FMR measurements. $\theta_H$ defines the angle between the external magnetic field, $H_{ext}$, and a direction of an electric current, $I_{RF}$ and $I_{DC}$. **c**, Temperature dependence of the device resistance. In order to distinguish the hysteretic effects, the resistance measured during heating (red) and cooling (blue) are indicated. **d**, ST-FMR field sweep spectra with 8 GHz, 3 dBm RF excitation at the sample temperatures of 290 K (blue), $(316 \pm 6)$ K (dark grey) and 355 K (red).

The relationship between the resonance linewidth $W$ and the driving frequency $f$ can be described by

$$W = W_0 + \frac{2\pi\alpha}{|\gamma|}f \qquad (2)$$

where $W_0$ is the inhomogeneous broadening, $\gamma$ is the electronic gyromagnetic ratio of NiFe (where a value of 185 GHz/T is used based on the electron Landé *g*-factor of 2.1[31]) and $\alpha$ is the effective Gilbert damping constant of the bilayer. The temperature dependence of the effective Gilbert damping parameter $\alpha$, as seen in Figure 3, is obtained by performing the ST-FMR measurements at different temperatures. The values of $\alpha$ increased from $(0.035 \pm 0.010)$ at 290 K to $(0.06 \pm 0.01)$ at 355 K. The increase of $\alpha$ with temperature can be attributed to the combined effects of the increased magnon population in NiFe at higher temperatures and the enhanced spin current absorption[32] at the VO$_2$/NiFe interface as the VO$_2$ becomes metallic (spin pumping enhancement). The total device resistance during the same thermal cycle is also plotted in the same figure to directly compare the temperature-dependence of $\alpha$ and the VO$_2$ phase transition.

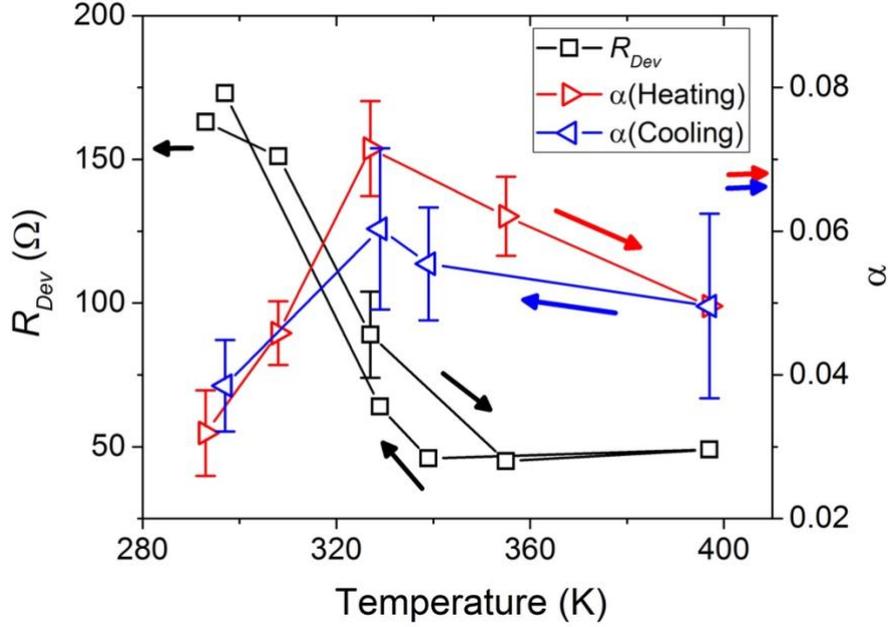

**Fig. 3: Temperature-depedence of device resistance and effective Gilbert damping parameter.** Effective Gilbert damping parameter, $\alpha$, of the $VO_2$/NiFe bilayer measured at different sample temperatures. To distinguish the hysteretic effects, data measured during heating (red right-tipped triangles) and cooling (blue left-tipped triangles) are indicated. The device resistance $R_{Dev}$ (black squares) is also plotted against temperature so that the changes in $\alpha$ can be compared directly the insulator-metal transition of $VO_2$.

## 3. TEMPERATURE-DEPENDENCE OF DC-INDUCED SPIN-ORBIT TORQUES

The next key step for functionalisation is to quantify current-induced spin-orbit torques in this system across the $VO_2$ phase transition. In order to examine this, we pass an additional DC current, $I_{DC}$, through the strip (as seen in the schematic in Figure 2b) and study how this additional stimulus affects the ST-FMR spectra, namely the linewidth $W$, at different temperatures. With the additional DC current, the current-induced spin-orbit torques generated in the $VO_2$ layer and the $VO_2$/NiFe interface exert damping-like torques on the precessing NiFe magnetisation. Depending on the sign of the torque (which in turn depends on the directions of the current, $I_{DC}$, and the external magnetic field, $H_{ext}$), the linewidth $W$ broadens or narrows. This can be observed in Figure 4a, where the application of the 1 mA (-1 mA) DC bias broadened (narrowed) the linewidth $W$ from the spectrum at 0 mA. The obtained changes in the linewidth, $\Delta W$, with our range of DC currents are a fraction of the original linewidth and comparable to the uncertainty in the linewidth estimation from the fitting according to the Equation (1). Therefore, the field sweep measurements are repeated between 20 and 50 times and the value of $W$ are obtained by averaging. The uncertainties in $W_{avg}$ are obtained by the statistical deviation of $W$, $\sigma W_{avg} = \frac{\sqrt{\sum_{i=1}^{N}(W_i - W_{avg})}}{N}$. The Table 1 shows the values of $W_{avg}$, $\sigma W_{avg}$, the number of field-sweeps and the average $R^2$ values for the measurements at 290 K.

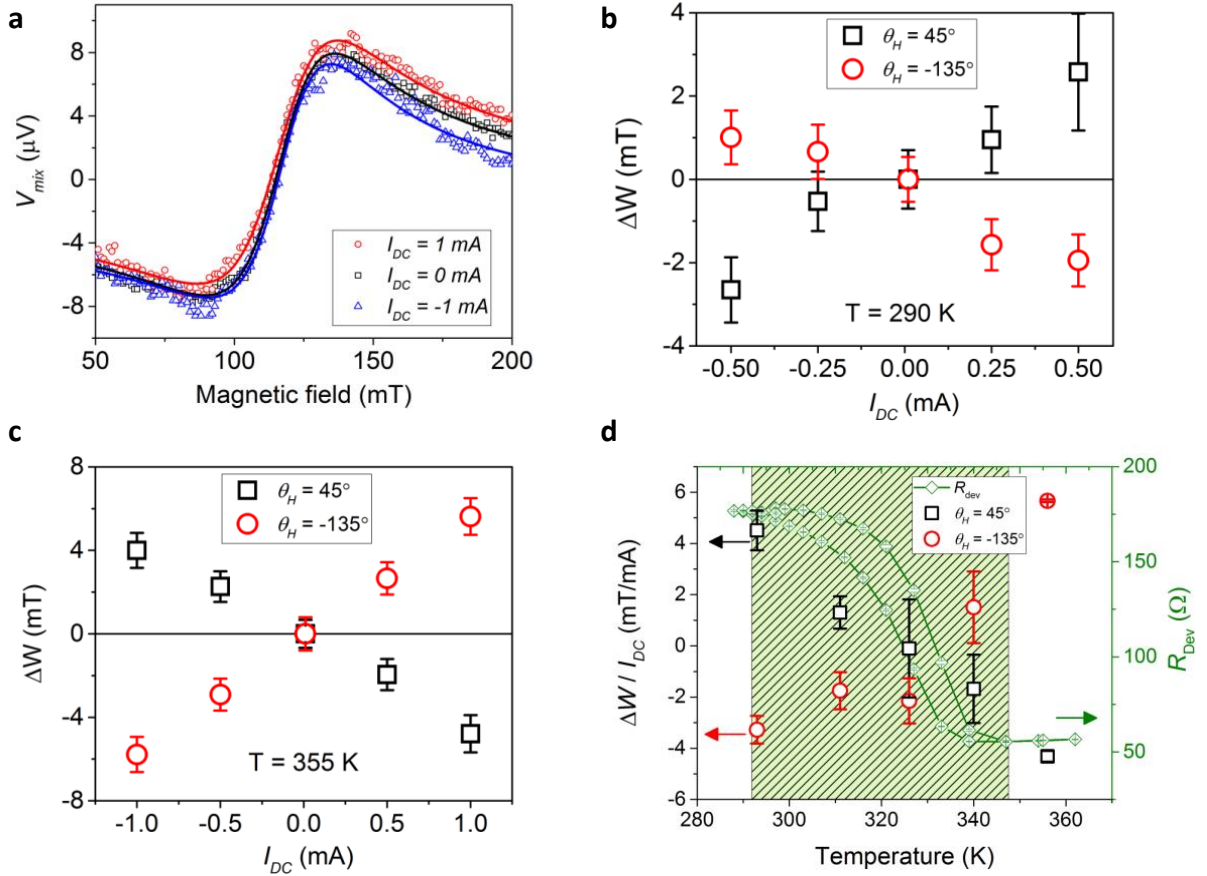

**Figure 4: DC- and temperature-dependence of the linewidth modulation. a**, ST-FMR field sweep data at 290 K with 8 GHz, 5dBm RF excitation at $\theta_H = 45°$ with the additional DC current, $I_{DC}$, of 0 mA (black), 1 mA (red) and -1 mA (blue). The solid lines are fits to the data using the Eq. (1). **b,c**, DC-current induced linewidth change, $\Delta W$, at 290 K (**b**) and 355 K (**c**) for $\theta_H = 45°$ (black squares) and $\theta_H = -135°$ (red circles). The lines are linear fits to the data points in each field directions. The total device resistances at 290 K and 355 K are 145 $\Omega$ and 59 $\Omega$, respectively. **d**, Temperature dependence of $\Delta W/I_{DC}$ for $\theta_H = 45°$ (black square) and -

135° (red circles), as compared to the device resistance (green diamonds). The light green shaded area indicates the temperature range of the insulator-metal transition.

**Table 1**: The average linewidths, errors, the number of field sweeps and the average $R^2$ values for the 290 K data.

| DC [mA] | $W_{avg,45°}$ [mT] | $\sigma W_{avg,45°}$ [mT] | $N$ | Average $R^2$ | $W_{avg,-135°}$ [mT] | $\sigma W_{avg,-135°}$ [mT] | $N$ | Average $R^2$ |
|---|---|---|---|---|---|---|---|---|
| **-0.5** | 25.2 | 0.4 | 48 | 0.94 | 27.9 | 0.4 | 49 | 0.96 |
| **-0.25** | 27.1 | 0.4 | 49 | 0.95 | 27.5 | 0.4 | 49 | 0.95 |
| **0.01** | 27.6 | 0.3 | 50 | 0.96 | 26.9 | 0.3 | 50 | 0.97 |
| **0.25** | 28.6 | 0.4 | 50 | 0.95 | 25.3 | 0.3 | 50 | 0.95 |
| **0.5** | 30.2 | 1.1 | 50 | 0.96 | 24.9 | 0.4 | 50 | 0.95 |

The changes in the linewidth $W$ with the added DC current is shown in Figure 4b and Figure 4c at 290 K and 355 K, respectively. (The ST-FMR spectra with DC currents at different temperatures can be seen in Supporting Information Figure S4.) The change in the linewidth is linearly proportional to the magnitude of the DC bias, indicating that the generated spin current is also linearly proportional to the applied charge current.

Remarkably, we observed a sign change of the torques across the phase transition of $VO_2$, suggesting competing origins of the spin-orbit torques. Figure 4d summarises the DC-induced linewidth changes, $\Delta W/I_{DC}$, at different temperatures, as compared to the device resistance across the phase transition. At 290 K, the $VO_2$ layer resistance is several orders of magnitude higher than that of the NiFe layer, and most of the applied DC current flows through the NiFe layer. The lack of the DC current flowing through the $VO_2$ layer eliminates the bulk spin Hall effect in the $VO_2$ as the main origin of the large spin-orbit torque observed at this temperature. The effect of the self-induced torque in NiFe, as observed in Ni[19] can explain our observed sign of the signal. Additional interfacial effects, such as inverse spin galvanic effect prominent in many Rashba-like interfaces[15,17,33], can also additionally contribute but these effects have been reported not to have a unique sign of the generated torques. Furthermore, as the interface between the $VO_2$ and the NiFe is present in both the low and the high temperature phase, it is not clear that strongly different inverse spin galvanic effects can be expected as a function of temperature.

As the temperature increases, $VO_2$ undergoes an insulator-metal transition and more current flows through the $VO_2$ layer. (The device resistance dependence of the $\Delta W/I_{DC}$ can be found in Supporting Information Figure S5.) Therefore, this charge current can create spin currents in the $VO_2$ layer by the bulk spin Hall effect, which competes with the other contributions such as the self-induced torque from NiFe or the interfacial torque. As seen in Figure 4d, the observed total spin-orbit torque decreases in magnitude with increasing temperature from 290 K, goes through the sign change at ~ 325 K near the middle of the insulator-metal transition then increases again in magnitude to 355 K. The spin-orbit torque generated via the bulk spin Hall effect in the metallic $VO_2$ layer at 355 K is of the same sign as seen in V/CoFeB[34] and $VO_2$/YIG[32] (negative effective spin Hall angle).

## 4. FIRST-PRINCIPLES CALCULATION OF SPIN HALL CONDUCTIVITY IN METALLIC $VO_2$ & DICSUSSION

Taking into account all the above points, the large changes in the spin-orbit torque observed in our system can be interpreted by two competing mechanisms. One of the major changes brought forward by the insulator-metal transition is the electric current flowing within the $VO_2$ layer. In the metallic phase of the $VO_2$, this current in turn generates spin/orbital Hall current, which is injected into NiFe layer and thus exerts a torque. In order to estimate this effect quantitatively, we performed first-principles calculations of spin and orbital Hall conductivities of the metallic $VO_2$ in the rutile structure, as seen in Figure 5. In the figure, we show $\sigma_{SH}$ (blue solid line) and $\sigma_{OH}$ (red dashed line) as a function of the Fermi energy ($E_F$) with respect to the true Fermi energy ($E_F^{true}$), where $E_F$ is varied assuming that the potential is fixed to the potential for $E_F^{true}$. The result indicates that there are two peaks for $\sigma_{SH}$ near $E_F \approx E_F^{true}$. On the other hand, a peak of $\sigma_{OH}$ is located ~ 0.3 eV above $E_F^{true}$. The values for the spin and orbital Hall conductivities at the true Fermi energy are $\sigma_{SH} = -96\ (\hbar/e)(\Omega \cdot cm)^{-1}$ and $\sigma_{OH} = +320\ (\hbar/e)(\Omega \cdot cm)^{-1}$, respectively. More details of the calculation can be found in Supporting Information (Section IV). Although the orbital Hall conductivity is larger than the spin Hall conductivity, its contribution to the torque is expected to be negligible here since the orbital-to-spin conversion ratio in NiFe is expected to be less than 10%[20,35]. We would like to point out that the sign of computed spin Hall conductivity is consistent with the sign of the effective spin Hall angle measured in the experiment, which allows us to conclude that the spin

Hall effect of the VO₂ is one of the main mechanisms for the torque when VO₂ is driven into the metallic phase. Meanwhile, there can be another contribution by the so-called self-induced torque/anomalous spin-orbit torque[19] in the ferromagnetic layer itself. As predicted and experimentally observed previously[19,36], this can be interpreted as the transfer of spin angular momentum between spin-polarised charge currents and magnetisation. While the spin Hall conductivity from this anomalous spin-orbit torque in NiFe itself was found to be large at ~2,300 S/cm, the value reduces to 10 – 100 S/cm at an interface with a non-magnetic layer such as Cu and AlOₓ, due to the additional angular momentum loss to the lattice via spin-orbit coupling. The expected magnitude and the positive sign of the spin Hall conductivity explains well the observed behaviour in our NiFe/VO₂ bilayer system across the VO₂ phase transition. At the VO₂ insulating regime the spin-orbit torque arises purely from the self-induced torque of the NiFe layer, while as the VO₂ becomes more metallic across the phase transition, the bulk and the negative spin Hall effect from the VO₂ dominates and reverses the spin-orbit torque direction.

Finally, there may be a contribution to the torque originating in the interfacial scattering, but we expect that the interfacial contribution does not change drastically across the insulator-metal transition since the strain induced by the structural phase transition is small (typically of ~ 1%[27]).

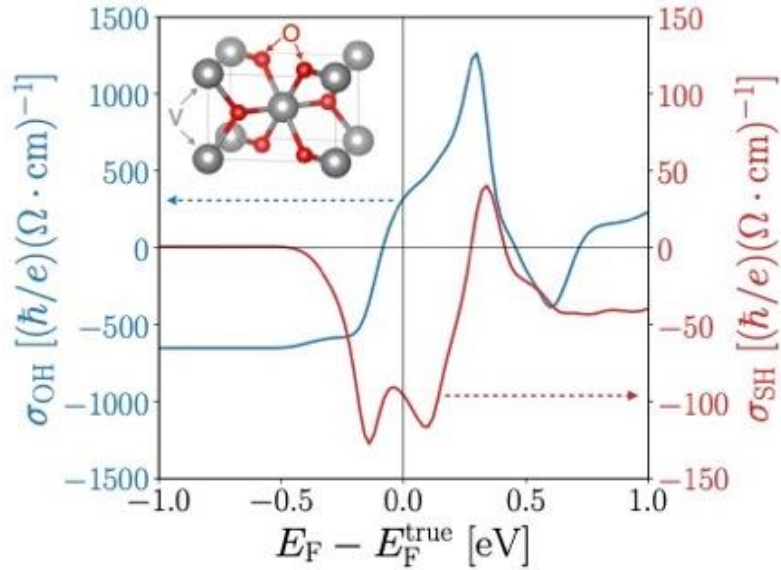

**Figure 5: Spin and orbital Hall conductivity for the metallic VO₂.** The crystal structure is shown in upper left, where grey and red spheres represent V and O atoms, respectively. The spin Hall conductivity ($\sigma_{\text{SH}}$) and orbital Hall conductivity ($\sigma_{\text{OH}}$) are displayed as a function of the Fermi energy ($E_{\text{F}}$) with respect to the true Fermi energy ($E_{\text{F}}^{\text{true}}$), which are indicated by solid blue and red lines, respectively. The values at $E_{\mathrm{F}}^{\mathrm{true}}$ are $\sigma_{\mathrm{SH}} = -96\ (\hbar/e)(\Omega \cdot \mathrm{cm})^{-1}$ and $\sigma_{\mathrm{OH}} = +320\ (\hbar/e)(\Omega \cdot \mathrm{cm})^{-1}$, respectively.

We can now compare our results with the previous spin-pumping inverse spin Hall effect (SP-ISHE) measurements in $VO_2$/YIG[32]. In this system, the only source of the observed ISHE signal is the spin-to-charge conversion within the $VO_2$. Therefore, there is no phase-dependent reversal of the signal with temperature, but only the broadening and the reduction of the signal due to the increased interface spin-transparency at the high temperature metallic phase, which is also observed in our case as the increase in the Gilbert damping parameter $\alpha$ (Figure 3). In $VO_2$/YIG, the SP-ISHE signal is largely affected by the conductivity change of the $VO_2$, which is observed as a sharp decrease in the signal at high temperature. In our system, the ST-FMR signal depends on the rectified AMR effect in NiFe, whose conductivity does not change significantly across the $VO_2$ phase transition. This allows the measurements of spin-orbit torques present at the $VO_2$/ferromagnet interface directly across the $VO_2$ phase transition.

As studied in depth using X-ray absorption spectroscopy[22], the change in the $VO_2$ orbital occupation across the phase transition is likely to affect directly the current-induced spin-orbit torque generation mechanisms. The investigation of the orbital correlation and its effect in the spin-orbit torque at the $VO_2$/ferromagnet interface is reserved for a future work.

## 5. CONCLUSIONS

We have measured the current-induced spin-orbit torques in the $VO_2$/NiFe bilayer system using the spin-torque ferromagnetic resonance technique. A sign change of the damping-like spin-orbit torques with temperature is observed across the $VO_2$ layer phase transition. The sign change and the modulation of the observed torques with temperature suggest coexistence of various competing mechanisms, mainly the bulk spin Hall effect in metallic $VO_2$, corroborated by our first-principles calculation, and the self-induced torque in NiFe. While additional interfacial effects can play a role, we expect these not to change significantly across the transition, but additional measurements could be carried out to identify further possible contributions. For applications, the large ($\pm100\%$) modulation, as well as the sign change of the spin-orbit torque enables full tunability of the torque to any desired value via device thermal history engineering, leading to drastically different device architectures.

## Supporting Information

Supporting Information is available from the Wiley Online Library or from the author.

## Acknowledgements

This is a highly interactive project. The experiments were conceived jointly, and the VO$_2$ samples were synthesized and characterized at University of California San Diego (UCSD). The magnetic stack deposition and the ST-FMR measurements were performed at Johannes Gutenberg University Mainz (JGU Mainz). The manuscript was written through multiple iterations by the authors. The research at UCSD was supported by the Office of Basic Energy Science, U.S. Department of Energy, BES-DMS and funded by the Department of Energy's Office of Basic Energy Science, DMR under grant DE FG02 87ER-45332. The work at JGU Mainz was funded by the Deutsche Forschungsgemeinshaft (DFG, German Research Foundation) – TRR 173 – 268565370 (projects A01, A11 and B02) and we acknowledge financial support from the Horizon 2020 Framework Programme of the European Commission under FET-Open grant agreement no. 863155 (s-Nebula) and ERC synergy grant agreement no. 856538 (3DMAGiC). We also gratefully acknowledge the Jülich Supercomputing Centre and RWTH Aachen University for providing computational resources under projects jiff40 and jara0062. D.-S. Han acknowledges the support from the National Research Foundation of Korea (NRF) funded by the Ministry of Science and ICT (2020R1C1C1012664, 2019M3F3A1A02071509) and the National Research Council of Science & Technology (NST) (No. CAP-16-01-KIST). K. Lee also acknowledges the support from Korea University Grant (K2111401).

## Conflict of Interest

The authors declare no conflict of interest.

## Author contributions

M.K., I.K.S. and D.-S.H. proposed and designed the experiments. P.S., P.N.L. and N.M.V. optimized, deposited and characterised the VO$_2$ samples. J.C. deposited the magnetic stack by sputtering. J.K. fabricated the devices. D.-S.H., J.K. and K.L. constructed the ST-FMR measurement system. J.K. performed the ST-FMR experiments and analysed the experimental data with help from G.K. and M.K. D.G and Y.M. performed the first-principles spin Hall conductivity calculations. J.K. wrote the manuscript through multiple iterations by all authors.

[REFERENCES]

(1)   Prinz, G. A. Magnetoelectronics. *Science* **1998**, *282* (5394), 1660–1663. https://doi.org/10.1126/science.282.5394.1660.

(2)   Davidson, A.; Amin, V. P.; Aljuaid, W. S.; Haney, P. M.; Fan, X. Perspectives of Electrically Generated Spin Currents in Ferromagnetic Materials. *Phys. Lett. A* **2020**, *384* (11), 126228. https://doi.org/10.1016/j.physleta.2019.126228.

(3)   Wunderlich, J.; Kaestner, B.; Sinova, J.; Jungwirth, T. Experimental Observation of the Spin-Hall Effect in a Two-Dimensional Spin-Orbit Coupled Semiconductor System. *Phys. Rev. Lett.* **2005**, *94* (4), 047204. https://doi.org/10.1103/PhysRevLett.94.047204.

(4)   Lou, X.; Adelmann, C.; Crooker, S. A.; Garlid, E. S.; Zhang, J.; Reddy, K. S. M.; Flexner, S. D.; Palmstrøm, C. J.; Crowell, P. A. Electrical Detection of Spin Transport in Lateral Ferromagnet–Semiconductor Devices. *Nat. Phys.* **2007**, *3* (3), 197–202. https://doi.org/10.1038/nphys543.

(5)   Liu, L.; Pai, C.-F.; Li, Y.; Tseng, H. W.; Ralph, D. C.; Buhrman, R. A. Spin-Torque Switching with the Giant Spin Hall Effect of Tantalum. *Science* **2012**, *336* (6081), 555–558. https://doi.org/10.1126/science.1218197.

(6)   Manchon, A.; Železný, J.; Miron, I. M.; Jungwirth, T.; Sinova, J.; Thiaville, A.; Garello, K.; Gambardella, P. Current-Induced Spin-Orbit Torques in Ferromagnetic and Antiferromagnetic Systems. *Rev. Mod. Phys.* **2019**, *91* (3), 035004. https://doi.org/10.1103/RevModPhys.91.035004.

(7)   Miron, I. M.; Garello, K.; Gaudin, G.; Zermatten, P.-J.; Costache, M. V.; Auffret, S.; Bandiera, S.; Rodmacq, B.; Schuhl, A.; Gambardella, P. Perpendicular Switching of a Single Ferromagnetic Layer Induced by In-Plane Current Injection. *Nature* **2011**, *476* (7359), 189–193. https://doi.org/10.1038/nature10309.

(8)   Garello, K.; Avci, C. O.; Miron, I. M.; Baumgartner, M.; Ghosh, A.; Auffret, S.; Boulle, O.; Gaudin, G.; Gambardella, P. Ultrafast Magnetization Switching by Spin-Orbit Torques. *Appl. Phys. Lett.* **2014**, *105* (21), 212402. https://doi.org/10.1063/1.4902443.

(9)   Hirohata, A.; Yamada, K.; Nakatani, Y.; Prejbeanu, I.-L.; Diény, B.; Pirro, P.; Hillebrands, B. Review on Spintronics: Principles and Device Applications. *J. Magn. Magn. Mater.* **2020**, *509*, 166711. https://doi.org/10.1016/j.jmmm.2020.166711.

(10)  Kato, Y. K. Observation of the Spin Hall Effect in Semiconductors. *Science* **2004**, *306* (5703), 1910–1913. https://doi.org/10.1126/science.1105514.

(11)  Sinova, J.; Valenzuela, S. O.; Wunderlich, J.; Back, C. H.; Jungwirth, T. Spin Hall Effects. *Rev. Mod. Phys.* **2015**, *87* (4), 1213–1260. https://doi.org/10.1103/RevModPhys.87.1213.

(12)  Sánchez, J. C. R.; Vila, L.; Desfonds, G.; Gambarelli, S.; Attané, J. P.; De Teresa, J. M.; Magén, C.; Fert, A. Spin-to-Charge Conversion Using Rashba Coupling at the Interface between Non-Magnetic Materials. *Nat. Commun.* **2013**, *4* (1), 2944. https://doi.org/10.1038/ncomms3944.

(13)  Ganichev, S. D.; Ivchenko, E. L.; Bel'kov, V. V.; Tarasenko, S. A.; Sollinger, M.; Weiss, D.; Wegscheider, W.; Prettl, W. Spin-Galvanic Effect. *Nature* **2002**, *417* (6885), 153–156. https://doi.org/10.1038/417153a.

(14)  Skinner, T. D.; Olejník, K.; Cunningham, L. K.; Kurebayashi, H.; Campion, R. P.; Gallagher, B. L.; Jungwirth, T.; Ferguson, A. J. Complementary Spin-Hall and Inverse Spin-Galvanic Effect Torques in a Ferromagnet/Semiconductor Bilayer. *Nat. Commun.* **2015**, *6*. https://doi.org/10.1038/ncomms7730.

(15)  Lesne, E.; Fu, Y.; Oyarzun, S.; Rojas-Sánchez, J. C.; Vaz, D. C.; Naganuma, H.; Sicoli, G.; Attané, J.-P.; Jamet, M.; Jacquet, E.; George, J.-M.; Barthélémy, A.; Jaffrès, H.; Fert, A.; Bibes, M.; Vila, L. Highly Efficient and Tunable Spin-to-Charge

Conversion through Rashba Coupling at Oxide Interfaces. *Nat. Mater.* **2016**, *15* (12), 1261–1266. https://doi.org/10.1038/nmat4726.

(16) Mellnik, A. R.; Lee, J. S.; Richardella, A.; Grab, J. L.; Mintun, P. J.; Fischer, M. H.; Vaezi, A.; Manchon, A.; Kim, E.-A.; Samarth, N.; Ralph, D. C. Spin-Transfer Torque Generated by a Topological Insulator. *Nature* **2014**, *511* (7510), 449–451. https://doi.org/10.1038/nature13534.

(17) DC, M.; Grassi, R.; Chen, J.-Y.; Jamali, M.; Reifsnyder Hickey, D.; Zhang, D.; Zhao, Z.; Li, H.; Quarterman, P.; Lv, Y.; Li, M.; Manchon, A.; Mkhoyan, K. A.; Low, T.; Wang, J.-P. Room-Temperature High Spin–Orbit Torque Due to Quantum Confinement in Sputtered BixSe(1–x) Films. *Nat. Mater.* **2018**, *17* (9), 800–807. https://doi.org/10.1038/s41563-018-0136-z.

(18) Rojas-Sánchez, J.-C.; Oyarzún, S.; Fu, Y.; Marty, A.; Vergnaud, C.; Gambarelli, S.; Vila, L.; Jamet, M.; Ohtsubo, Y.; Taleb-Ibrahimi, A.; Le Fèvre, P.; Bertran, F.; Reyren, N.; George, J.-M.; Fert, A. Spin to Charge Conversion at Room Temperature by Spin Pumping into a New Type of Topological Insulator: α -Sn Films. *Phys. Rev. Lett.* **2016**, *116* (9), 096602. https://doi.org/10.1103/PhysRevLett.116.096602.

(19) Wang, W.; Wang, T.; Amin, V. P.; Wang, Y.; Radhakrishnan, A.; Davidson, A.; Allen, S. R.; Silva, T. J.; Ohldag, H.; Balzar, D.; Zink, B. L.; Haney, P. M.; Xiao, J. Q.; Cahill, D. G.; Lorenz, V. O.; Fan, X. Anomalous Spin–Orbit Torques in Magnetic Single-Layer Films. *Nat. Nanotechnol.* **2019**, *14* (9), 819–824. https://doi.org/10.1038/s41565-019-0504-0.

(20) Go, D.; Lee, H.-W. Orbital Torque: Torque Generation by Orbital Current Injection. *Phys. Rev. Res.* **2020**, *2* (1), 013177. https://doi.org/10.1103/PhysRevResearch.2.013177.

(21) Filianina, M.; Hanke, J.-P.; Lee, K.; Han, D.-S.; Jaiswal, S.; Rajan, A.; Jakob, G.; Mokrousov, Y.; Kläui, M. Electric-Field Control of Spin-Orbit Torques in Perpendicularly Magnetized W / CoFeB / MgO Films. *Phys. Rev. Lett.* **2020**, *124* (21), 217701. https://doi.org/10.1103/PhysRevLett.124.217701.

(22) Aetukuri, N. B.; Gray, A. X.; Drouard, M.; Cossale, M.; Gao, L.; Reid, A. H.; Kukreja, R.; Ohldag, H.; Jenkins, C. A.; Arenholz, E.; Roche, K. P.; Dürr, H. A.; Samant, M. G.; Parkin, S. S. P. Control of the Metal–Insulator Transition in Vanadium Dioxide by Modifying Orbital Occupancy. *Nat. Phys.* **2013**, *9* (10), 661–666. https://doi.org/10.1038/nphys2733.

(23) del Valle, J.; Salev, P.; Tesler, F.; Vargas, N. M.; Kalcheim, Y.; Wang, P.; Trastoy, J.; Lee, M.-H.; Kassabian, G.; Ramírez, J. G.; Rozenberg, M. J.; Schuller, I. K. Subthreshold Firing in Mott Nanodevices. *Nature* **2019**, *569* (7756), 388–392. https://doi.org/10.1038/s41586-019-1159-6.

(24) Luo, H.; Wang, B.; Wang, E.; Wang, X.; Sun, Y.; Li, Q.; Fan, S.; Cheng, C.; Liu, K. Phase-Transition Modulated, High-Performance Dual-Mode Photodetectors Based on WSe $_2$/VO $_2$ Heterojunctions. *Appl. Phys. Rev.* **2019**, *6* (4), 041407. https://doi.org/10.1063/1.5124672.

(25) Zylberstejn, A.; Mott, N. F. Metal-Insulator Transition in Vanadium Dioxide. *Phys. Rev. B* **1975**, *11* (11), 4383–4395. https://doi.org/10.1103/PhysRevB.11.4383.

(26) Wei, G.; Lin, X.; Si, Z.; Lei, N.; Chen, Y.; Eimer, S.; Zhao, W. Phase-Transition-Induced Large Magnetic Anisotropy Change in VO2/(Co/Pt)2 Heterostructure. *Appl. Phys. Lett.* **2019**, *114* (1), 012407. https://doi.org/10.1063/1.5058751.

(27) Wei, G.; Lin, X.; Si, Z.; Wang, D.; Wang, X.; Fan, X.; Deng, K.; Liu, K.; Jiang, K.; Lei, N.; Chen, Y.; Mangin, S.; Fullerton, E.; Zhao, W. Optically Induced Phase Change for Magnetoresistance Modulation. *Adv. Quantum Technol.* **2020**, *3* (3), 1900104. https://doi.org/10.1002/qute.201900104.

(28)   Fan, X.; Wei, G.; Lin, X.; Wang, X.; Si, Z.; Zhang, X.; Shao, Q.; Mangin, S.; Fullerton, E.; Jiang, L.; Zhao, W. Reversible Switching of Interlayer Exchange Coupling through Atomically Thin VO2 via Electronic State Modulation. *Matter* **2020**, *2* (6), 1582–1593. https://doi.org/10.1016/j.matt.2020.04.001.

(29)   He, J.; Di Sante, D.; Li, R.; Chen, X.-Q.; Rondinelli, J. M.; Franchini, C. Tunable Metal-Insulator Transition, Rashba Effect and Weyl Fermions in a Relativistic Charge-Ordered Ferroelectric Oxide. *Nat. Commun.* **2018**, *9* (1), 492. https://doi.org/10.1038/s41467-017-02814-4.

(30)   Liu, L.; Moriyama, T.; Ralph, D. C.; Buhrman, R. A. Spin-Torque Ferromagnetic Resonance Induced by the Spin Hall Effect. *Phys. Rev. Lett.* **2011**, *106* (3), 036601. https://doi.org/10.1103/PhysRevLett.106.036601.

(31)   Nan, T.; Emori, S.; Boone, C. T.; Wang, X.; Oxholm, T. M.; Jones, J. G.; Howe, B. M.; Brown, G. J.; Sun, N. X. Comparison of Spin-Orbit Torques and Spin Pumping across NiFe/Pt and NiFe/Cu/Pt Interfaces. *Phys. Rev. B* **2015**, *91* (21), 214416. https://doi.org/10.1103/PhysRevB.91.214416.

(32)   Safi, T. S.; Zhang, P.; Fan, Y.; Guo, Z.; Han, J.; Rosenberg, E. R.; Ross, C.; Tserkovnyak, Y.; Liu, L. Variable Spin-Charge Conversion across Metal-Insulator Transition. *Nat. Commun.* **2020**, *11* (1), 476. https://doi.org/10.1038/s41467-020-14388-9.

(33)   Current-Induced Torques and Interfacial Spin-Orbit Coupling. *Phys. Rev. B 88* (21), 214417. https://doi.org/10.1103/PhysRevB.88.214417.

(34)   Wang, T.; Wang, W.; Xie, Y.; Warsi, M. A.; Wu, J.; Chen, Y.; Lorenz, V. O.; Fan, X.; Xiao, J. Q. Large Spin Hall Angle in Vanadium Film. *Sci. Rep.* **2017**, *7* (1), 1306. https://doi.org/10.1038/s41598-017-01112-9.

(35)   Go, D.; Freimuth, F.; Hanke, J.-P.; Xue, F.; Gomonay, O.; Lee, K.-J.; Blügel, S.; Haney, P. M.; Lee, H.-W.; Mokrousov, Y. Theory of Current-Induced Angular Momentum Transfer Dynamics in Spin-Orbit Coupled Systems. *Phys. Rev. Res.* **2020**, *2* (3), 033401. https://doi.org/10.1103/PhysRevResearch.2.033401.

(36)   Amin, V. P.; Li, J.; Stiles, M. D.; Haney, P. M. Intrinsic Spin Currents in Ferromagnets. *Phys. Rev. B* **2019**, *99* (22), 220405. https://doi.org/10.1103/PhysRevB.99.220405.